\documentclass[prl,cha, preprintnumbers ,amsmath,amssymb, showpacs, raggedbottom, floatfix]{revtex4}

\usepackage{bm}
\usepackage[colorlinks=true,linkcolor=blue]{hyperref}
\usepackage{amsmath}
\usepackage{amssymb}
\usepackage{amsthm}
\usepackage{amsfonts}
\usepackage{times}
\usepackage{enumerate}
\usepackage{latexsym}
\usepackage{ifpdf}
\usepackage{graphicx}
\usepackage{color}
\usepackage{makeidx}
\usepackage{ulem}

\expandafter\ifx\csname package@font\endcsname\relax\else
\expandafter\expandafter
\expandafter\usepackage
\expandafter\expandafter
\expandafter{\csname package@font\endcsname}%
\fi
\begin{document}

\title{Understanding Magnetism in Double Double Perovskites: A Complex Multiple Magnetic Sublattice System}
\author{Anita Halder$^{1,2}$}
\author{Shreya Das$^1$}
\author{Prabuddha Sanyal$^3$}
\author{Tanusri Saha-Dasgupta$^{1}$}
\email{t.sahadasgupta@gmail.com}

\address{$^1$Department of Condensed Matter Physics and Material Sciences,
S.N. Bose National Centre for Basic Sciences, JD Block, Sector-III, Salt 
Lake City, Kolkata 700 106, India}
\address{$^2$ School of Physics, Trinity College Dublin, Dublin, Ireland.}
\address{$^3$ Maulana Abul Kalam Azad University of Technology, Kolkata, India.} 
\maketitle

\textbf{Understanding magnetism in multiple magnetic sublattice system, driven by the interplay of varied nature
  of magnetic exchanges, is on one hand challenging and on other hand intriguing.
  Motivated by the recent synthesis of AA$^{'}$BB$^{'}$O$_6$ double double perovskites with multiple magnetic
  ions both at A- and B-sites, we investigate the mechanism of magnetic behavior in these interesting class of
  compounds. We find that the magnetism in such multiple sublattice compounds is governed by the interplay and delicate 
  balance between two distinct mechanisms, a) kinetic energy-driven multiple sublattice double exchange mechanism and
  b) the conventional super-exchange mechanism. The derived spin Hamiltonian based on first-principles calculations
is solved by classical Monte Carlo technique which reproduces the observed magnetic properties. Finally, the
influence of off-stoichiometry, as in experimental samples, is discussed. Some of these double double perovskite compounds
are found to possess large total magnetic moment and also are found to be half-metallic, which raises the hope of future applications of these large magnetic moment half-metallic oxides in spintronics and memory devices.} 
  
\bigskip

Perovskite structured ABO$_3$ transition metal oxides remained the holy grail of condensed matter physics due to wide range of fascinating properties exhibited by them, which includes properties like high temperature superconductivity, colossal magneto-resistance, half-metallicity etc.\cite{C. N. R Rao(1989), AS Bhalla(2000)} With an aim to tailor properties further, one of the common route is cation substitution. Substitution and 1:1 ordering of cations in B sublattices in rock-salt arrangement give rise to A$_2$BB$^{'}$O$_6$ double perovskites.\cite{King (2010), Tanusri(2013), Sami Vasala(2015), Tanusri(2020)} The topic of magnetism in transition metal oxides with two magnetic ions as in double perovskite structure has received
significant attention.\cite{D.D. Sarma(2000), K.-I. Kobayashi(1998), Hena Das(2008), DP1, DP2, DP3, DP4, Prabuddha Sanyal(2009), kato(2007),Krockenberger(2007),Nyrissa S. Rogado(2005),Hena Das(2009),Prabuddha Sanyal(2017),Hena Das(2011),Anita(2019),kartik(2015)}

In this backdrop, it is an interesting issue to ask, what happens if the compounds involve even larger number of magnetic ions, {\it i.e.} more than two magnetic ions, as in double double perovskites. Due to the structural and compositional flexibility of the structure, perovskites can accommodate almost all of the elements of the periodic table, and also can support various possible coordination. In recent time double double perovskites of general formula
AA$^{'}_{0.5}$A$^{''}_{0.5}$BB$^{'}$O$_6$ have been synthesized using high pressure and temperature,\cite{E. Solana-Madruga(2016), McNally (2017), E. Solana-Madruga(2018)} combining columnar ordering at A sublattice and rock-salt ordering at B sublattice with five independent cation sites, A, A$^{'}$, A$^{''}$, B and B$^{'}$, hosting rare-earth or alkaline-earth ion at A site, 3d transition metals at A$^{'}$, A$^{``}$ and B sites, and 5d transition metal at B$^{'}$ site. Use of high pressure is able to stabilize small magnetic transition-metal ions such as Mn$^{2+}$ at the A-sites of perovskites in place of large, nonmagnetic cations like Ca$^{2+}$ and Sr$^{2+}$, with reduced coordination of tetrahedral (4) and square planar (4) instead of usual dodecahedral (12) coordination of A site.\cite{ A. J. Dos santos-Garca(2015)} This introduces a source of magnetism at A-site, which in turn drives the interplay of magnetism between multiple sublattices, resulting in highly enriched magnetic properties.

One would naively expect presence of multiple magnetic ions with multiple magnetic exchange would lead to a frustrating situation and spin glass like ground state. Contrary to this expectation, recently synthesized double double perovskites CaMnMReO$_6$ (M=Ni,Co) are found to be magnetically ordered, \cite{Elena Solana-Madruga(2019)} showing
ferromagnetic ordering in CaMnNiReO$_6$ with parallel alignment of spins, which is found to change to
ferrimagnetic ordering when Ni is replaced by Co, the neighboring element in periodic table. We note the net moment of such multi-component ferromagnetic system is very high, paving the way to design large moment magnetic oxides.
This makes the situation rather curious in the sense, what makes the three or more magnetic sublattice system CaMnNiReO$_6$ ferromagnetic, and why replacement of Ni by Co, the neighboring element in periodic table,  makes it ferrimagnetic.  What is the driving mechanism of magnetism in such multi sublattice magnetic system ? Understanding of such a complex, multiple magnetic sublattice system is expected to bring out rich physics, which would help future designing of such oxides.

Motivated by these developments, we present here a first-principles density functional theory (DFT) based study of these compounds which takes into account the structural and chemical details in an accurate way, followed by construction
of DFT-derived spin Hamiltonian, which is solved with Monte Carlo (MC) simulation. Our study uncovers a novel
exchange mechanism to be operative in these compounds, which turn out to be a combination of multi-sublattice
hybridization or kinetic energy-driven double-exchange mechanism, and the more conventional super-exchange mechanism,
the nature of ground state magnetic order being decided by the competition of these two. While for CaMnNiReO$_6$, the
multi-sublattice hybridization-driven double-exchange mechanism wins over the super-exchange mechanism stabilizing
the long range ferromagnetic state, the replacement of Ni by Co, increases
the core spin value at B sublattice from S=1 to S=3/2, thus toggling the balance between
two exchange mechanisms, favoring the long range ferrimagnetic behavior. 
The spin-Hamiltonian with parameters derived in a first-principles
manner provide good description of measured magnetic properties.\cite{Elena Solana-Madruga(2019)} Introduction of
off-stoichiometry is found to maintain the magnetic ground state, encouraging exploration of many more candidates in
such multiple magnetic sublattice double perovskites. Interestingly, the large magnetic moment compounds, arising due to
long range ferromagnetic ordering between multiple magnetic sublattices, also turned out to
be half-metallic having important implication for spintronic applications.

\section{Results}

\subsection{Crystal Structure}

Fig. 1 shows the four formula unit tetragonal, P42/n crystal structure of stoichiometric CaMnNiReO$_6$ (CMNRO).\cite{Elena Solana-Madruga(2019)} CaMnCoReO$_6$ (CMCRO) compound is isostructural to CaMnNiReO$_6$. The structure consists of four magnetic sublattices, tetrahedrally coordinated 3d transition metal (TM) Mn1 at A$^{'}$ site, square planar coordinated 3d transition metal Mn2 at A$^{``}$ site, octahedrally coordinated 3d transition metal Ni/Co at B site and octahedrally coordinated 5d transition metal Re at B$^{'}$ site, making it a 3d-5d TM magnetic system. Mn1 is connected to two nearest neighbour (NN) Mn2 sites through Mn1-O-O-Mn2 superexchange paths, while it is connected to 4 NN Ni(Co)/Re through Mn1-O-Ni(Co)/Re super-exchange paths. Ni(Co) and Re are connected to each other through corner shared Ni(Co)-O-Re, of bond angles 141-152$^{o}$.

\subsection{Electronic Structure}

We analyze the electronic structure of the studied compounds in terms of spin-polarized density of states, and
its projection to orbital characters which provide us the information on charge and spin states of the transition metal
ions. The GGA+$U$ density of states (DOS), with choice of $U$ = 5 (2) eV and $J_H$ = 0.9 (0.4) eV at 3d TM (Re) sites, for CMNRO and CMCRO are shown in top and bottom panels of Fig. 2, respectively.
In agreement with experimental findings, the ground state of CMNRO is found to be ferromagnetic, with moments at three 3d TM sublattices Mn1, Mn2 and Ni
sites aligned in parallel direction, while the moment at Re site is found to be aligned opposite to the moments at Mn1, Mn2
and Ni sites. The calculated magnetic moments at Mn (Mn1 and Mn2), Ni and Re are found to be 4.5 $\mu_B$, 1.6 $\mu_B$
and 0.5 $\mu_B$ respectively, with a large total moment of 24 $\mu_B$ in the unit cell. On the contrary the ground state of CMCRO is
found to be ferrimagnetic, as observed experimentally, with moments of Mn1, and Mn2 aligned in antiparallel direction,
and Co moment pointing in the direction of Mn1. The Re moment is found to be antiparallel to Mn1 and Co, with calculated moments values of 4.5 $\mu_B$
(Mn1 and Mn2), 2.6 $\mu_B$ (Co) and 0.5 $\mu_B$ (Re) and total moment of 8 $\mu_B$ in the unit cell. The calculated moments are in
conformity with nominal 2+ valence of Mn1 and Mn2 with high spin \textit{d$^5$} occupancy, 2+ valence of Ni/Co with high spin (HS) \textit{d$^8$}/\textit{d$^7$} occupancy and 6+ valence of Re with \textit{d$^1$} occupancy. Following this, in DOS of CMNRO we find Mn1 and Mn2 states are filled in the majority spin channel and empty in the minority channel. Ni \textit{e$_g$} DOS in CMNRO get filled in majority spin channel and empty in minority, while Ni \textit{t$_{2g}$} states are filled in both spin channels. The partially filled Re \textit{t$_{2g}$} states in CMNRO, with one electron in the minority spin channel and strongly hybridized with Mn1/Mn2 \textit{d} and Ni \textit{e$_g$} states, crosses the Fermi level making the solution metallic in minority spin channel and gaped in the majority spin channel. This half metallic solution persists in CMCRO, though Mn1 and Mn2 \textit{d} states now become filled and empty, respectively in two opposite spin channels and Co \textit{t$_{2g}$} becomes partly empty. This points to possibility of achieving spin-dependent nature of the carrier scattering in these compounds, with a large spin value, which would allow for the resistance of these large moment compounds to be strongly influenced by the low magnetic field.

Effect of spin-orbit coupling (SOC) was checked, which is expected to be appreciable for 5d TM element, Re. The qualitative results are found to remain unchanged upon inclusion of SOC, apart from an appreciable orbital moment of $\sim$ 0.15 $\mu_B$ that develops at Re site, antiparallel to its spin moment.

\subsection{Mechanism of Magnetism}


In order to shed light on the mechanism of magnetism in this interesting class of compounds, we derive the low energy spin Hamiltonian out of DFT inputs. For
this purpose, we perform muffin tin orbital based downfolding calculations\cite{O. K. Andersen(2000)} that integrate out degrees of freedom which are not of interest in an energy selective manner.  Wannier representation of the downfolded Hamiltonian provides the estimates of the onsite energies and the hopping interactions between the orbitals retained in the basis during the process of downfolding. In the first step of downfolding calculations, we retain the Mn1, Mn2 \textit{d} states, Ni \textit{e$_g$}/ Co \textit{d} states and Re \textit{t$_{2g}$} in the basis and integrate out the rest. In second step, Mn1, Mn2 and Ni/Co degrees of freedom are downfolded retaining only the Re \textit{t$_{2g}$} degrees of freedom in the basis. The latter massive downfolding provides the estimates of the Re \textit{t$_{2g}$} onsite energies renormalized by the hybridization from Mn1, Mn2 and Ni/Co states. Thus the onsite matrix elements of the real space Hamiltonian defined in the first and second step of downfolding
calculations, give the energy level positions before and after switching on the hybridization between Mn1/Mn2/Ni(Co) and Re states, respectively. Results of two step downfolding calculations for CMNRO and CMCRO are presented in top and bottom panels of Fig. 3, respectively. Mn1 -\textit{d}, Mn2-\textit{d}, Ni \textit{e$_g$} (Co -\textit{d} ) and Re \textit{t$_{2g}$} states are both crystal field split and exchange split. In distorted tetrahedral coordination, Mn1 \textit{d} states are split into 1-1-1-2 fold degeneracies, while Mn2 \textit{d} states in square planar coordination are split into 2-2-1 fold degeneracies. The trigonal distortion in ReO$_6$ octahedra splits its \textit{t$_{2g}$} states in 1-2 fold degeneracies.

Examination of Fig. 3 reveals several interesting aspects, key to construct the low energy spin Hamiltonian. First of all, in Mn1-Mn2-Ni(Co)-Re basis, Re \textit{t$_{2g}$} states are essentially non-magnetic with negligible exchange splitting. Secondly, the Re \textit{t$_{2g}$} states lie within the exchange split states of Mn1 \textit{d}, Mn2 \textit{d} and Ni \textit{e$_g$}/Co \textit{d}. Third and most importantly, upon switching on the hybridization between Mn1 \textit{d}/Mn2 \textit{d}/Ni \textit{e$_g$}(Co \textit{d}) and Re \textit{t$_{2g}$}, captured through massive downfolding procedure, an exchange splitting of 0.6-0.8 eV is induced among Re \textit{t$_{2g}$} states, with the direction of spin splitting opposite to that of Mn1 or Mn2 or Ni/Co. This essentially establishes the hybridization-driven multi sublattice double-exchange process to be operative, in which a negative spin splitting in the essentially non-magnetic site is induced through hybridization between the localized spin and itinerant electrons.\cite{D.D. Sarma(2000),K.-I. Kobayashi(1998)} In the present context, this can be captured in terms of a 3+1 sublattice Kondo Lattice model, consisting of
a) a large core spin at the Mn1, Mn2 and Ni(Co) sites, b) strong coupling on the Mn1/Mn2/ Ni (Co) site between the core spin and the itinerant electron, strongly preferring one spin polarization of the itinerant electron, and c) delocalization of the itinerant electron on the Mn1-Mn2-Ni(Co)-Re network, in the similar spirit as in \cite{Prabuddha Sanyal(2009)},

\begin{eqnarray}
H_{DE}&=& \epsilon_{B}\sum_{i\sigma}b_{i\sigma}^{\dagger}b_{i\sigma}+\epsilon_{Mn1}\sum_{i}m^{1\dagger}_{i\sigma}m^{1}_{i\sigma}  \nonumber \\
&+&\epsilon_{Mn2}\sum_{i}m^{2\dagger}_{i\sigma}m^{2}_{i\sigma}+\epsilon_{Re}\sum_{i}r^{\dagger}_{i\sigma}r_{i\sigma} \nonumber \\
&+& t_{B-Re}\sum_{<ij>}(b_{i\sigma}^{\dagger}r_{j\sigma}+ h.c.) \nonumber \\
&+&t_{Mn1-Re}\sum_{<ij>}(m_{i\sigma}^{1\dagger}r_{j\sigma}+h.c.) \nonumber \\
&+&t_{Mn2-Re}\sum_{<ij>}(m_{i\sigma}^{2\dagger}r_{j\sigma}+h.c.) \nonumber \\
&+&J_{B}\sum_{i\in B} \vec{S}_{i}^{B}\cdot b_{i\alpha}^{\dagger}\vec{\sigma}_{\alpha\beta}b_{i\beta} \nonumber \\
&+&J_{Mn1}\sum_{i\in A^{\prime}} \vec{S}_{i}^{Mn1}\cdot m_{i\alpha}^{1\dagger}\vec{\sigma}_{\alpha\beta}m_{i\beta}^{1} \nonumber \\
&+&J_{Mn2}\sum_{i\in A^{\prime\prime}} \vec{S}_{i}^{Mn2}\cdot m_{i\alpha}^{2\dagger}\vec{\sigma}_{\alpha\beta}m_{i\beta}^{2} \nonumber \\
\end{eqnarray}

The $m$'s refer to the Mn sites and the $b$'s to the B (Ni/Co) sites. $t_{B-Re}$, $t_{Mn1-Re}$, $t_{Mn2-Re}$ represent the nearest neighbor B-Re, Mn1-Re, Mn2-Re
  hoppings respectively, with onsite elements $\epsilon_{B}$, $\epsilon_{Mn1}$, $\epsilon_{Mn2}$ and $\epsilon_{Re}$.
The ${\bf S}_i$ are `classical' (large $S$)
core spins at the Mn1/Mn2/B sites, coupled to the itinerant Re electrons through a coupling $J$, when the Re electron hops onto the respective
sublattice.

It is to be noted that in $H_{DE}$, the Kondo coupling parameter $J$ is present only at the
magnetic Mn1, Mn2 and B sites, which possess a large core spin (S=5/2 at Mn1 and Mn2, and S=1 for Ni and S=3/2 for Co),
with which the itinerant Re electron interacts when it hops onto these magnetic sites. The $J/W$ ratio between the Kondo exchange
coupling, $J$ and the bandwidth, $W$ is thus relevant only on the magnetic Mn1, Mn2 and B (Ni/Co) sites. Following the DFT inputs, the
  calculated $J/W$ ratios for the Mn1, Mn2 and Ni sites in CMNRO are found to be 3.77, 2.06 and 2.7 respectively, while
  the ratios for the Mn1, Mn2 and Co sites in CMCRO are given by 3.529, 1.87 and 3.5 respectively. Thus the exchange
  coupling $J$ is appreciably larger than the bandwidth $W$ for all the magnetic sites. Moreover, since the bandwidth,
  $W \approx z \times t$ where $t$'s are the hopping parameters appearing in the Hamiltonian, and $z$ is the number of neighbors,
  the $J/t$ ratios are even larger, of the order of 8 or 10, justifying use of $J \rightarrow$ $\infty$ model. The $J \rightarrow$ $\infty$ limit of double exchange models was used by Anderson and Hasegawa\cite{Anderson-hasegawa} and
later studied by P. De Gennes\cite{Gennes}  in the context of perovskites. This limit was studied in the context of
double perovskites in Ref.\onlinecite{DP3}. $J \rightarrow$ $\infty$ approximation has been used for other magnetic perovskite and
  double perovskite compounds in the literature for similar $J/t$ ratios.\cite{manganites, millis, Prabuddha Sanyal(2009), DP3}

Invoking the $J \rightarrow$ $\infty$ approximation, one can derive the effective spin Hamiltonian for CMNRO in terms of the core spins at Mn1 (S=5/2), Mn2 (S=5/2)
and Ni (S=1) site as given in the following. The details of the derivation can be found in the supplementary information (SI).

\begin{eqnarray} 
H^{'}_{DE} & = & 4D_{Mn1-Mn2}\sum_{<ij>i\in A',j\in A''}\sqrt{\frac{1+\mathbf{S}^{Mn1}_{i}\cdot\mathbf{S}^{Mn2}_{j}}{2}} \nonumber \\
  &+ & 8D_{Mn1-Ni}\sum_{<ij> i\in A',j\in B}\sqrt{\frac{1+\mathbf{S}^{Mn1}_{i}\cdot\mathbf{S}^{Ni}_{j}}{2}}  \nonumber \\
&+&8D_{Mn2-Ni}\sum_{<ij>i\in A'',j\in B}\sqrt{\frac{1+\mathbf{S}^{Mn2}_{i}\cdot\mathbf{S}^{Ni}_{j}}{2}} 
\end{eqnarray}

 A similar Hamiltonian can be written for the Co compound, with $\mathbf{S}^{Ni}$ (S=1) replaced by $\mathbf{S}^{Co}$ (S=3/2), where the coupling constants are $D_{Mn1-Mn2}$, $D_{Mn1-Co}$ and $D_{Mn2-Co}$. 

The above described multi sublattice double-exchange Hamiltonian although is capable of describing the ferromagnetic state of CMNRO, does not account for the fact that replacement of Ni by Co in CMCRO changes ferromagnetic state to ferrimagnetic state. This suggests that together with H$_{DE}$ another source of magnetism needs to be considered. Indeed, there  exists another source of magnetism, namely the super-exchange between the half-filled Mn1-\textit{d}, Mn2-\textit{d}, Ni \textit{e$_g$}, and high spin Co (\textit{t$_{2g}$+e$_{g}$}) states.
The Goodenough-Kanamori rule\cite{Goodenough} states that superexchange interactions are antiferromagnetic where the virtual electron transfer is between overlapping orbitals that are each half-filled, but they are ferromagnetic where the virtual electron transfer is from a half-filled to an empty orbital or from a filled to a half-filled orbital. Following this, the super-exchange contributions are all antiferromagnetic in nature (cf Fig. 4) with its strength defined by hopping integrals ($t$) and onsite energy differences ($\Delta$), $J$ $\propto$ $\sum_{m,m^{'}} t_{m,m^{'}}^2/$ (U+ $\Delta_{m,m^{'}}$), where $m$ and $m^{'}$ are the orbitals at site $i$ (Mn1/Mn2/Ni (Co)) and $j$ (Mn1/Mn2/Ni (Co)). 

The net Hamiltonian for CMNRO adding the contribution of double exchange and super-exchange is thus given by,

\begin{eqnarray}
  H & = & 4D_{Mn1-Mn2}\sum_{<ij>i\in A',j\in A''}\sqrt{\frac{1+\mathbf{S}^{Mn1}_{i}\cdot\mathbf{S}^{Mn2}_{j}}{2}} \nonumber \\
  & + & 8D_{Mn1-Ni}\sum_{<ij>i\in A',j\in B}\sqrt{\frac{1+\mathbf{S}^{Mn1}_{i}\cdot\mathbf{S}^{Ni}_{j}}{2}}\nonumber \\
&+& 8D_{Mn2-Ni}\sum_{<ij>i\in A'',j\in B}\sqrt{\frac{1+\mathbf{S}^{Mn2}_{i}\cdot\mathbf{S}^{Ni}_{j}}{2}} \nonumber \\
  &  + & 4J_{Mn1-Mn2}\sum_{<ij>i\in A',j\in A''}\mathbf{S}^{Mn1}_{i}\cdot\mathbf{S}^{Mn2}_{j} \nonumber \\
  & + & 8J_{Mn1-Ni}\sum_{<ij>i\in A',j\in B}\mathbf{S}^{Mn1}_{i}\cdot\mathbf{S}^{Ni}_{j} \nonumber \\ 
&+&8J_{Mn2-Ni}\sum_{<ij>i\in A'',j\in B}\mathbf{S}^{Mn2}_{i}\cdot\mathbf{S}^{Ni}_{j}
\end{eqnarray}

and a similar one for CMCRO.\\

In order to estimate the various coupling constants, $D_{Mn1-Mn2}$, $D_{Mn1-Ni/Co}$, $D_{Mn2-Ni/Co}$ and $J_{Mn1-Mn2}$, $J_{Mn1-Ni/Co}$ and $J_{Mn2-Ni/Co}$, we apply a two step process. In the first step, we apply downfolding procedure \cite{O. K. Andersen(2000)} to construct a spin unpolarized Mn1-Mn2-Ni(Co) Hamiltonian defined in effective Mn1-\textit{d}, Mn2-\textit{d}, Ni \textit{e$_g$} (Co \textit{d}) basis. The real space representative of this Hamiltonian provides the estimate of onsite matrix elements of  Mn1-\textit{d}, Mn2-\textit{d}, Ni \textit{e$_g$} (Co \textit{d}) and the hopping interactions (assumed to be nearest neighbour) between them. Following this, $J_{Mn1-Mn2}$, $J_{Mn1-Ni/Co}$ and $J_{Mn2-Ni/Co}$ were estimated using the super-exchange formula, $J$  =
$\varSigma_{m,m^{'}}$ 2 $\times$ $t_{m,m^{'}}^2/$ (U+ $\Delta_{m,m^{'}}$).
In the second step, the total energies for different possible spin configurations at Mn1, Mn2 and Ni(Co) sites are calculated, and mapped on to the spin Hamiltonian given in Eqn. 2. Putting the values of $J_{Mn1-Mn2}$, $J_{Mn1-Ni/Co}$ and $J_{Mn2-Ni/Co}$ obtained from super-exchange formula, the estimates of $D_{Mn1-Mn2}$, $D_{Mn1-Ni/Co}$ and $D_{Mn2-Ni/Co}$ are obtained. The estimated values of $D$'s, and $J$'s, for the two compounds are given in Table I.

For CMNRO, we find that effective Mn1-Mn2, Mn1-Ni and Mn2-Ni interactions are all negatively signed, {\it i.e} ferromagnetic, in conformity with the ferromagnetic ground state found in the experiment, as well as in DFT total energy calculations. Similarly for CMCRO, we find Mn1-Mn2 and Mn2-Co effective interactions are positively signed {\it i.e} antiferromagnetic, while Mn1-Co interaction is ferromagnetic in conformity with its ferrimagnetic ground state.

Inspecting Table I, we further find while the strength of Mn1-Mn2 super-exchange ($J_{Mn1-Mn2}$) remains similar between the two compounds, the strength of Mn1/Mn2 - B super-exchange is greatly enhanced in CMCRO compared to CMNRO, $J_{Mn1-Ni/Co}$ being enhanced by a factor of 1.6 and $J_{Mn2-Ni/Co}$ being enhanced by a factor of 3.2. This is expected due to the fact that while for Ni two unpaired \textit{e$_g$} electrons participate in the super-exchange process, for Co, three unpaired electrons, two belonging to \textit{e$_g$} manifold and one belonging to \textit{t$_{2g}$} contribute. At the same time, we find a significant weakening of the hybridization-driven exchange between Mn1-Mn2, reduced by two orders of magnitude compared to Ni compound. These important changes turn the net interaction to be antiferromagnetic between Mn1 and Mn2, and that between Mn2 and Co, compared to all ferro interaction for Ni compound.

\subsection{Monte Carlo Study of the Spin Hamiltonian}

In order to evaluate the finite temperature properties of the defined spin Hamiltonian, described by Eqn. 3, we perform
Monte Carlo simulations. The total energy of a particular spin configuration can be obtained from the spin Hamiltonian
by plugging in input parameters $D$'s and $J$'s, as listed in Table I.
The spin configurations at Mn1, Mn2 and Ni/Co sites are generated through Metropolis algorithm in
a 3$\times$3$\times$3 unit cell simulation box
with periodic boundary condition. Starting from an initial
temperature of 400 K (1000 K) for CMNRO (CMCRO) the simulation temperature is stepped down
to T = 1 K with an interval of 2 K. Hundred thousand Monte Carlo steps are employed to
ensure a large sample space, while the physical quantity like magnetization is
calculated by averaging over last 10,000 Monte Carlo steps.
The magnetizations plotted as a function of temperature for CMNRO and CMCRO are
shown in Fig. 5.  

For CMNRO compound, our Monte Carlo simulation correctly reproduces the ground state of this compound where the spins of Mn1. Mn2 and Ni are all found to be aligned in parallel to each other (cf top, left panel, Fig. 5).  We note that the total moment at low temperature is found out to be 28$\mu_B$/unit cell, corresponding to the sum of the nominal moment 5$\mu_B$ for 2 Mn1 and 2 Mn2 with the nominal moment 2$\mu_B$ for 4 Ni sites. Transition temperature (T$_c$) may be obtained from the inflection point of the derivative of magnetization versus temperature curve, as shown in the top panel, Fig 5. The T$_c$ is found to be 142 K which is close to experimentally reported value of 158 K.\cite{Elena Solana-Madruga(2019)} For CMCRO compound, the ferrimagnetic ground state is also correctly captured. In this case, Mn2 is found to be antiparallel to Mn1 and Co moment, giving rise to the total moment of
12 $\mu_B$/unit cell, arising from magnetic moment of
3$\mu_B$ in 4  Co sites and cancellation of moments at Mn1 and Mn2 sites. The transition in the case of CMNRO is
noticeably sharper compared to CMCRO, as reflected in the narrower width of the inverse peak in dM/dT curve. Additionally in CMCRO, a shoulder is observed in the left of the inverse peak which is completely absent in CMNRO.

By repeating the calculation with larger simulation cell size of 4$\times$4$\times$4 (see SI), we find
the peak and shoulder structure in CMCRO is robust, which arises due to the competition between effective
FM Mn1-Co interaction and the two effective AFM Mn1-Mn2 and Mn1-Co interactions. To demonstrate the effect of
competing nature of magnetic interactions in dM/dT curve of Co compound, we further present the dM/dT
curve for varying $D_{Mn1-Co}$ value in the inset of bottom right panel of Fig. 5. As found, upon reducing $D_{Mn1-Co}$
from DFT estimated value of -143.9 meV to -135.9 meV, {\it i.e.} weakening ferro interaction, the  high
temperature peak is shifted to lower temperature, along with redistribution of weight between the peak and
the shoulder, converting the shoulder to a peak. Thus the shoulder feature is reminiscent of second peak which
is not resolved when the strength of ferro and antiferro interactions are comparable. This implies the high
temperature peak feature arises from the ferro interaction, with the lower temperature feature arising due
to antiferro interaction. The experimental study\cite{Elena Solana-Madruga(2019)} on
CaMnReCoO$_6$ reports only magnetic susceptibility, and do not report dM/dT. However reported experimental dM/dT data
for multiple magnetic ion containing Nd$_2$NiMnO$_6$ double perovskite does exhibit such two feature structure.\cite{das}
Such two feature dM/dT curve is also seen for ferrimagnetic compound NiCr$_2$O$_4$.\cite{nicro}
We notice with choice of DFT estimated value of $D_{Mn1-Co}$, the second feature appears around 200 K, very
close to experimentally reported T$_C$ of 188 K.\cite{Elena Solana-Madruga(2019)}

 \subsection{Effect of Off-stoichiometry}

 The discussion above involves stoichiometric compounds, while the experimental samples of CMNRO and CMCRO are
 reported to be off-stoichiometric.\cite{Elena Solana-Madruga(2019)} The experimental samples show high degree of
 B-site cation ordering with nominal antisite disorder of 3.4 and 2.5$\%$ for Co and Ni compounds respectively.
 This has been attributed to high degree of charge contrast between (Mn/Co/Ni)$^{2+}$ and Re$^{6+}$.
 However, for CMCRO, while there is 96$\%$ Co at the octahedral B site,
 30-40$\%$ of Co was reported to substitute Mn at the A-sites, leading to an overall Co-rich composition of
 CaMn$_{0.7}$Co$_{1.3}$ReO$_6$ as opposed to the stoichiometric formula of CaMnCoReO$_6$. Similarly, for CMNRO, there
 is an overall Ni-poor composition of CaMn$_{1.2}$Ni$_{0.8}$ReO$_6$ in the experimental sample with some of the Mn atoms
 occupying Ni sites in B sublattice. We therefore, need to check whether the above theoretical understanding also holds
 good for the off-stoichiometric compounds.

 In order to mimic these off-stoichiometric situations, we replace one out of the four Ni atoms in
 the unit cell by Mn, giving rise to Ni poor composition CaMn$_{1.25}$Ni$_{0.75}$ReO$_6$, close to experimental
 composition. Since all four Ni sites are equivalent in the unit cell, any one chosen out of four possible sites, give rise
 to same results. Similarly, for CMCRO, an extra Co atom replacing one of the four Mn atoms in the unit cell is introduced,
 giving rise to composition CaMn$_{0.75}$Co$_{1.25}$ReO$_6$. Total energy calculations show that Co prefers to occupy
 the square planar Mn site (Mn2) over Mn1. Interestingly it is found that for CMNRO even in presence
 of off-stoichiometry the ground state remains ferromagnetic with Mn1, Mn2, Mn@Ni and Ni spins aligned
 in parallel. This highlights the dominant role of hybridization-driven magnetism, as opposed to super-exchange
 driven mechanism, which depends primarily on the positioning of energy levels, and not on the
 exchange pathways. Similarly, for CMCRO, even in presence of off-stoichiometry, Mn1 and Mn2 spins
 continue to remain antiparallel, while the Co spins (both at A$^{``}$ site and B site) are found to be oppositely
 aligned to Mn2. The magnetic ground states are thus found to be robust, and remain unaltered even in
 presence of off-stoichiometry, as also found experimentally. 

 The computed Mn1-Mn2, Mn1-Ni(Co), Mn2-Ni(Co) exchanges for CaMn$_{1.25}$Ni$_{0.75}$ReO$_6$ and CaMn$_{0.75}$Co$_{1.25}$ReO$_6$
 are found not to change significantly compared to their stoichiometric counterparts ($\sim$ 3-10$\%$) (see Table I), although
 introduction of off-stoichiometry introduces few additional interactions like in CMNRO, Mn@Ni-Mn1, Mn@Ni-Mn2 replacing some of
 Ni-Mn1, Ni-Mn2 interactions, respectively, and in CMCRO, Co@Mn2-Mn1, Co@Mn2-Co, replacing some of
 Mn2-Mn1, Mn2-Co interaction, respectively. Computation of these additional interactions show the signs
 of effective interactions corresponding to these additional interactions are the same as those of
 replacing interactions, with values within 5-7 $\%$.

 This suggests the magnetic transition temperature to be not altered drastically by the off-stoichiometry effect. To check this
 explicitly, we carried out Monte Carlo study of Ni-poor and Co-rich compounds as well. Within the 3$\times$3$\times$3 unit cell
 simulation box size, for the off-stoichiometric compounds, it is possible to have different possible atomic configurations of
 Mn@Ni and Co@Mn2. Total energy calculations show that extra Mn (Co) atoms at Ni (Mn2) sites prefer to be uniformly distributed
 rather than being clustered. Considering uniform distribution of extra Mn (Co) atoms in 3$\times$3$\times$3 unit simulation cell,
 this leads to 152 different configurations. To take this into account, the MC results are averaged over atomic configurations.

 In conformity with DFT results the ground state is found to be ferromagnetic for CMNRO and ferrimagnetic for CMCRO. The variation of moment
   with temperature results in two cases is shown
   Fig. 6. In presence of off-stoichiometry, the saturation moment for CMNRO and CMCRO becomes 31 $\mu_B$/unit cell and
   20 $\mu_B$/unit cell, respectively.
   The dM/dT curves presented in insets, show similarity with that found for stoichiometric compounds, confirming
   the transition temperature not be significantly effected by off stoichiometry.

\section{Summary and Outlook}

In this communication, we study the magnetism in systems containing multiple magnetic sublattices.
The study has been motivated by synthesis of double double perovskite compounds of general formula,
AA$^{'}_{0.5}$A$^{''}_{0.5}$BB$^{'}$O$_6$, having transition metal magnetic ions in both A and B sites.
The key findings of our study are summarized in the following,

\begin{description}
\item{$\bullet$} Our theoretical analysis combining first-principles and model Hamiltonian approaches, uncovers
  the microscopic origin of counter-intuitive long range ordered magnetism in
 double double perovskite compounds containing 3d magnetic ions at A and B sites, and 5d magnetic ions
  at B$^{'}$ sites, which turn out to be an interplay of hybridization-driven multi-sublattice double exchange
  and super-exchange mechanism of magnetism.

\item{$\bullet$}  This interplay relies on the positioning of the $d$ energy levels as well as the filling. This in turn,
  triggers ferromagnetic long range order in CMNRO compound containing two different Mn ions at A sites, and Ni and Re ions
  at B and B$^{'}$ sites, while replacement of Ni by Co at B site, decreasing the filling by one in CMCRO stabilizes ferrimagnetism,
  accounting for the experimental observations.\cite{Elena Solana-Madruga(2019)}

\item{$\bullet$} The spin Hamiltonian, capturing the interplay of hybridization-driven multi-sublattice double exchange
  and super-exchange mechanism of magnetism is parameterized in terms of three exchange constants
 $D_{Mn1-Mn2}$, $D_{Mn1-Ni/Co}$, $D_{Mn2-Ni/Co}$
  for the hybridization-driven multi-sublattice double exchange and another three exchange constants $J_{Mn1-Mn2}$, $J_{Mn1-Ni/Co}$, $J_{Mn2-Ni/Co}$ for the super-exchange, with the parameters derived from first-principles estimated hopping interactions, onsite energies
  and total energies of different spin configurations.

\item{$\bullet$} The computed temperature dependent magnetization by Monte Carlo reproduces the measured magnetic transition
  temperature of  CMNRO with reasonable accuracy. For CMCRO, it is found that the competition between ferro and antiferro nature
  of effective interactions, manifests as two hump structure of dM/dT, which should be probed further experimentally.

\item{$\bullet$} The calculations are further extended to off stoichiometric composition of Ni-poor and Co-rich
 compounds, in order to mimic the experimental situation. The magnetic properties are found to be retained even in presence
 of off-stoichiometry, since the hybridization-driven multi-sublattice double exchange, a dominant contributor in exchange
 mechanism of CMNRO and CMCRO, relies on energy level positioning, rather than on exchange pathways.

\end{description}

The proposed theory of magnetism being general in nature, should be applicable to multi sublattice mixed 3d-4d/5d
transition metal systems, where one of the transition metal element is a large band width 4d or 5d element with exchange
splitting significantly smaller than the band width. With appropriate choice of 3d and 4d/5d elements, this opens up the possibility of
stabilization of a large moment ferromagnetic state. Our first-principles calculation shows this ferromagnetic state with high value
of magnetization is furthermore half-metallic which should be an attractive possibility for spintronics applications.

 \section{Methods}

The first-principles DFT calculations are carried out using the plane-wave pseudo-potential method implemented
within the Vienna Ab-initio Simulation Package (VASP).\cite{G. Kresse(1996)} The exchange-correlation functional
is considered within the generalized gradient approximation (GGA).\cite{PBE} The projector-augmented wave (PAW)
potentials\cite{P. E. Blochl(1994)} are used and the wave functions are expanded in the plane-wave basis with a
kinetic-energy cut-off of 600 eV. Reciprocal-space integration is carried out with a $k$-space mesh of 6 $\times$
6 $\times$ 6. The exchange-correlation beyond GGA is treated within GGA+$U$ approach with local Coulomb interaction
parameterized in terms of Hubbard $U$ and Hund's coupling $J_H$ within the multi-band formulation.\cite{S. L. Dudarev(1998)} For double-counting correction, fully localized limit (FLL) of double-counting is considered since the around mean
field (AMF) double-counting is found to give magnetic states a significantly larger energy penalty
than that by the FLL counterpart.\cite{doublecounting} The parameters of GGA+$U$ calculations are chosen as
$U$ = 5 eV and $J_H$ = 0.9 eV for Ni/Co as appropriate for \textit{3d} TM atoms, and U = 2 eV and $J_H$ = 0.4 eV for Re as appropriate for 5d TM atoms.\cite{U-value} The U values are varied over 1-2 eV and the qualitative results
are found to remain unchanged.

In order to extract a few-band tight-binding (TB) Hamiltonian out of the full DFT calculation, $N$-th order muffin tin orbital $N$MTO) calculations are carried out.\cite{O. K. Andersen(2000)} A prominent feature of this method is the downfolding scheme.
Starting from a full DFT calculation, it defines a few-orbital Hamiltonian in an energy-selected, effective Wannier function basis,
by integrating out the degrees of freedom that are not of interest. The $N$-MTO technique relies on the self-consistent potential
parameters obtained out of linear muffin-tin orbital (LMTO)\cite{lmto} calculations.

The magnetization data, obtained from the Monte Carlo simulation of the spin Hamiltonian, are calculated on a N$\times$N$\times$N
unit cell of Mn and Ni/Co atoms. Finite size effect has been
checked. The results presented in the manuscript
are obtained from 3$\times$3$\times$3 lattice simulations. The periodic boundary conditions are applied during the simulation.
The magnetic transition temperatures are estimated from these calculations.


\section{Acknowledgments}
The authors acknowledge the support of DST Nano-mission for the computational
facility used in this study. T.S-D acknowledges J.C.Followship  (grant no. JCB/2020/000004) for research support.

\section{Author contributions statement}

A.H., S.D., P.S. did the theoretical calculations. A.H. and S.D. have equal contributions.
The figures were made by A.H. and S.D. The results were analyzed by T.S.-D, A.H., S.D., and P.S.
T.S.-D wrote the manuscript. A.H., S.D., P. S. and T.S.-D. finalized the manuscript.

\section{Competing financial interests}

The authors declare no competing financial interests.

\newpage

\begin{table}
\caption{Estimates of the coupling constants connecting the core spins for the double-exchange and super-exchange mechanisms operative in CMNRO  and CMCRO, as estimated employing the super-exchange formula and total energy calculations of different spin configurations. Negative (positive) signs indicate ferro-(antiferro-) magnetic interactions.}
  \begin{tabular}{cccc}
    \\\hline
\\    
    \textbf{CMNRO}	& $D$ (meV) & $JS^{2}$ (meV) & \textbf{Effective}\\
    \\
Mn1-Mn2		& -91.7 (-94.7)	& 47.5 (48.8)	& -44.2 (-45.9)\\
	    	&    		&   		& \\
Mn1-Ni      & -123.5 (-117.3)  & 54.8 (56.1)   & -68.7 (-61.2) \\
			&			&			& \\
Mn2-Ni		& -28.6 (-29.6)	& 10.8 (9.7)	& -17.8 (-19.9)\\
\\			&			&			&\\
Mn@Ni-Mn1      & -100.9  & 37.9  &            -63.0 \\ 
&			&			& \\
Mn@Ni-Mn2 &   -50.3  & 30.1  & -20.2  \\
&			&			& \\
\hline
\textbf{CMCRO}	& $D$ (meV) & $JS^{2}$ (meV) & \textbf{Effective}\\
\\
Mn1-Mn2		& 1.9 (2.1) 	& 48.5 (44.6)	& 50.4 ( 46.7)\\
	    	&    		&   		& \\
Mn1-Co          & -143.9 (-147.4)  & 87.9 (86.1)   & -56.0 (-61.3) \\
	    	&    		&   		& \\
Mn2-Co		& -18.7 (-19.8)	& 41.0 (43.1)	& 22.3 (23.3) \\
&    		&   		& \\
Co@Mn2-Mn1      & 2.4       & 47.7  &           50.1 \\ 
&			&			& \\
Co@Mn2-Co       &   -150.3  & 88.1  & -62.2  \\
&	&			& \\\hline
  \end{tabular}
  \end{table}

\begin{figure}
	\centering
	\includegraphics[width=0.9\columnwidth]{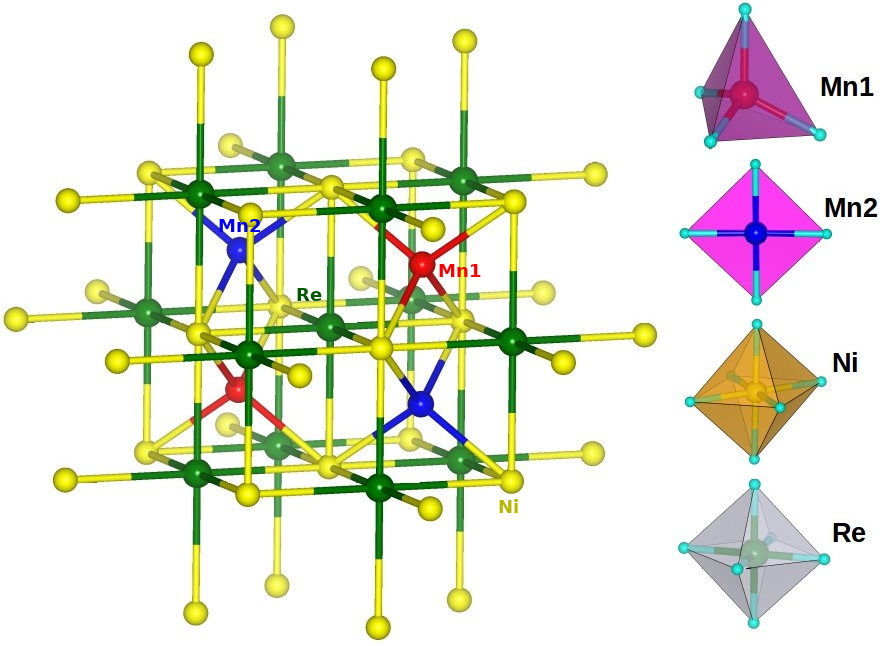}
	\caption{(Color online) Crystal structure of the stoichiometric CMNRO compound. CMCRO compound is isostructural to CMNRO. The left panel shows the three dimensional network of four magnetic ions in the structure with Mn at tetrahedral site (Mn1), Mn at square planar site (Mn2), Ni and Re atoms marked as red, blue, yellow and green coloured balls respectively.  The right panel shows the oxygen coordination of the four magnetic ions, tetrahedral for Mn1, square planar for Mn2, and octahedral for Ni and Re.}
\end{figure}

\begin{figure}
	\centering
	\includegraphics[width=0.7\columnwidth]{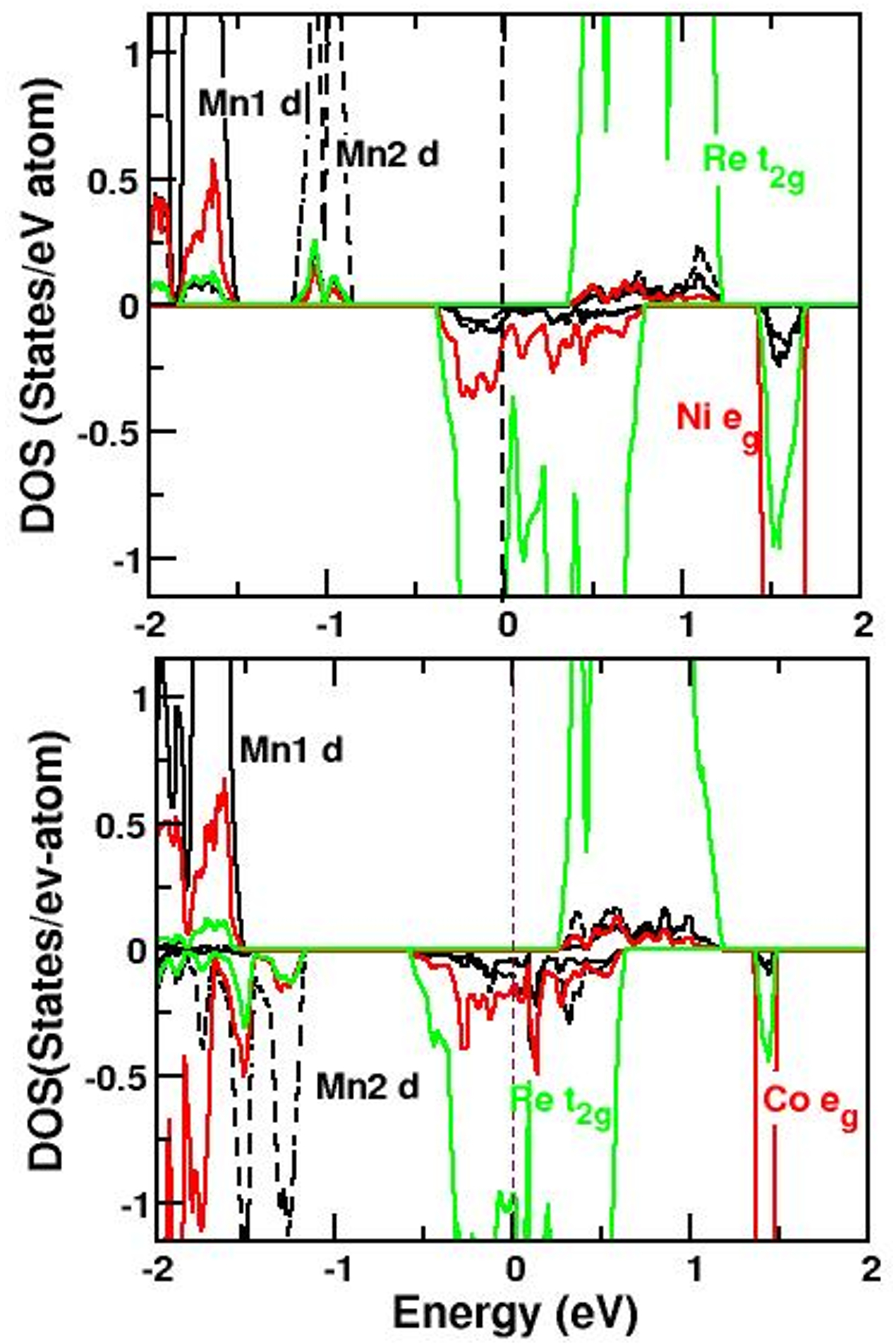}
	\caption{(Color online) The GGA+$U$ density of states of CMNRO (top) and CMCRO (bottom) projected to
            Mn1 \textit{d} (black, solid), Mn2 \textit{d} (black, dashed), Ni \textit{e$_g$}/ Co \textit{d}
            (red, solid) and Re \textit{t$_{2g}$} (green, solid) states. The zero of the energy is
            fixed at the Fermi energy.}
\end{figure}

\begin{figure}
	\centering
	\includegraphics[width=0.7\columnwidth]{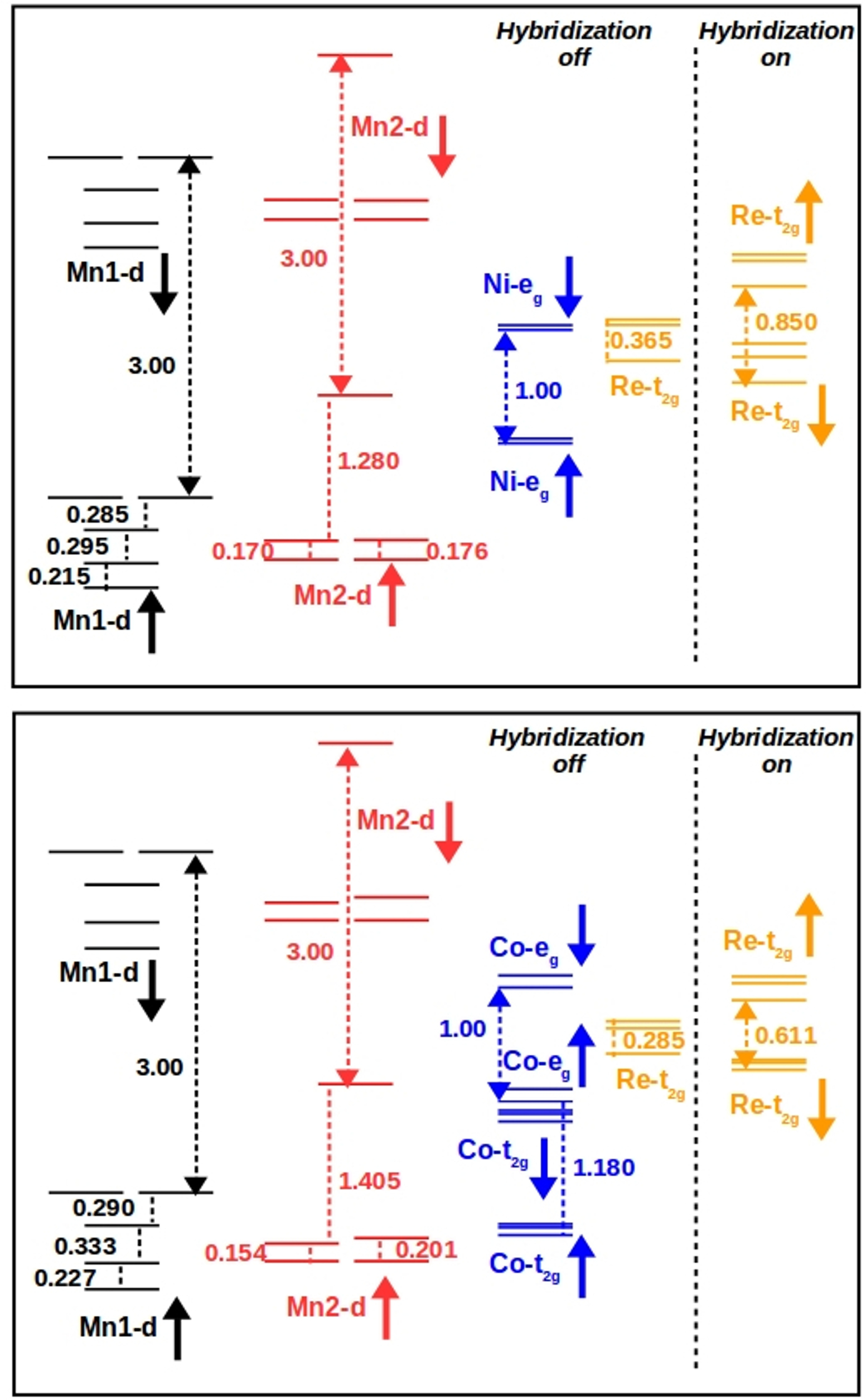}
	\caption{(Color online) The energy level diagram for CMNRO (top) and CMCRO (bottom) considering Mn1-\textit{d}/Mn2-\textit{d}/Ni \textit{e$_g$} (Co \textit{d})/Re \textit{t$_{2g}$} in basis (Hybridization-off) and in the massively downfolded Re \textit{t$_{2g}$} only basis (Hybridization-on). See text for details.}
\end{figure}

\begin{figure}
	\centering
	\includegraphics[width=\columnwidth]{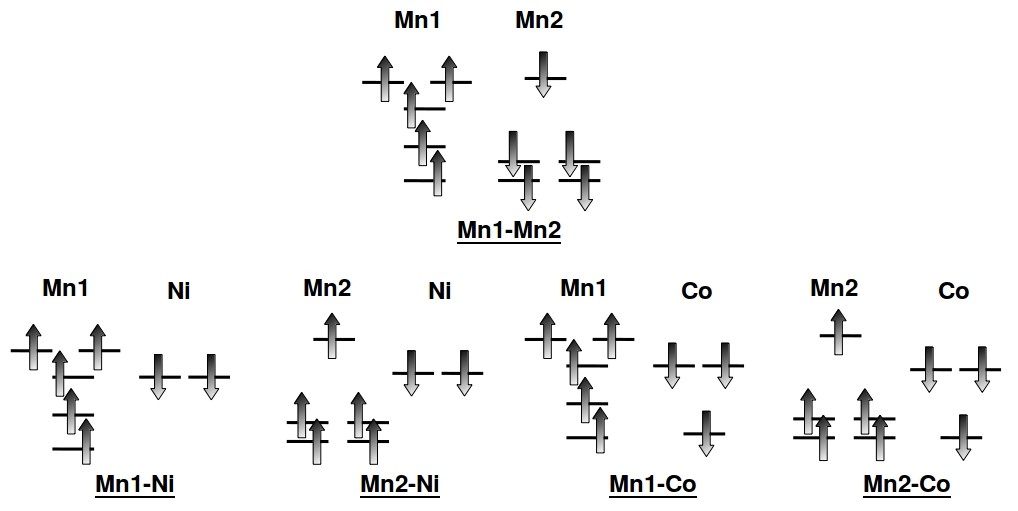}
	\caption{Super-exchange interactions in CMNRO and CMCRO between half-filled $d$ states of Mn1, Mn2, half-filled $e_g$ states of Ni and half-filled $e_g$ and one of the $t_{2g}$ states of Co. The fully filled levels not contributing in super-exchange are not shown.}
\end{figure} 

\begin{figure}
	\centering
	\includegraphics[width=\columnwidth]{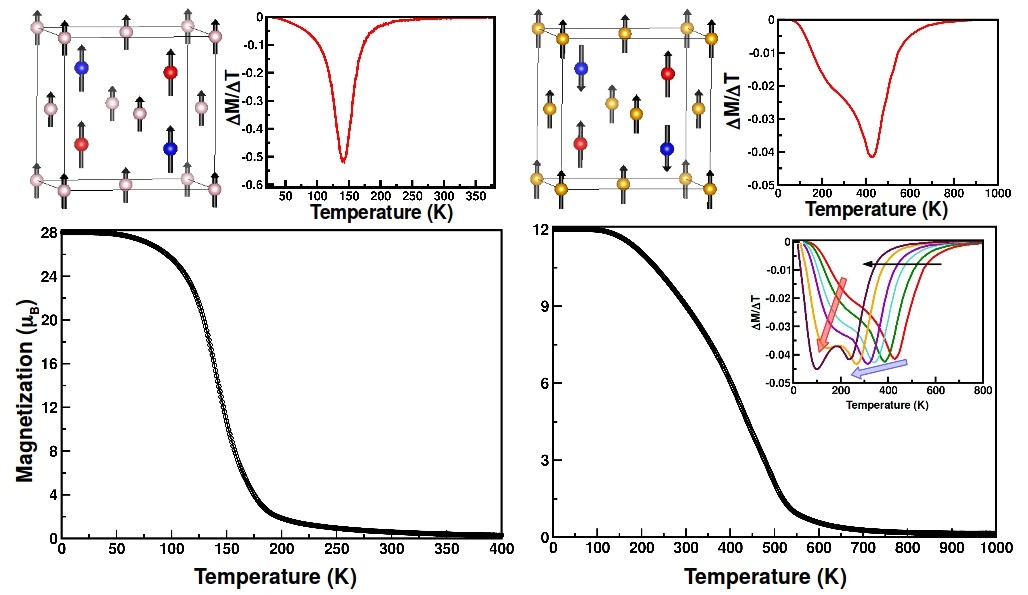}
	\caption{(Color online) Magnetic properties of CMNRO (left) and CMCRO (right) obtained from Monte-Carlo simulation. Top, left panels show the
          ground state magnetic structures, while the top, right panels show the derivative of magnetization as a function of temperature, the minimum corresponding to the transition temperature of the corresponding compound. Mn1, Mn2, Ni(Co) atoms are represented by red, blue and light pink (yellow) balls respectively. The lower panels shows the magnetization plotted as a function of temperature. The inset in lower, right panel shows the shift of transition temperature (the minima of the curves) for monotonic decrease of $D_{Mn1-Co}$. For details see text.}
\end{figure}

\begin{figure}
	\centering
	\includegraphics[width=\columnwidth]{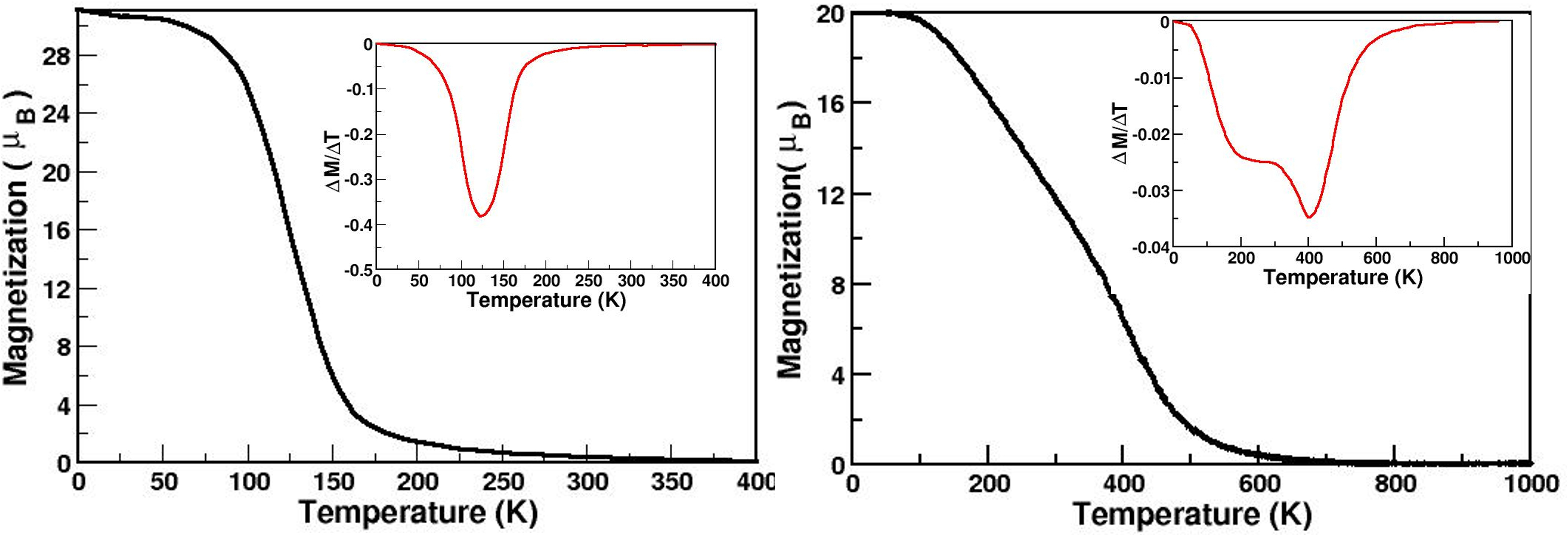}
	\caption{(Color online) Magnetic properties of CaMn$_{1.25}$Ni$_{0.75}$ReO$_6$ (left) and CaMn$_{0.75}$Co$_{1.25}$ReO$_6$  (right) obtained from
          Monte-Carlo simulation. Insets show the derivative of magnetization as a function of temperature.}
\end{figure}  


\begin{thebibliography}{999}
	\bibitem{C. N. R Rao(1989)}
	C. N. R. Rao {\em Annu. Rev. Phys. Chem.}, 
	{\bf 40}, 291 (1998).
	
	\bibitem{AS Bhalla(2000)}
	A. S. Bhall, R. Guo, R. Roy,{\em Mater Res Innov.} {\bf 4}, 3 (2000).
	
   \bibitem{King (2010)}
	G. King, P. M. Woodward, {\em J. Mater. Chem.}, {\bf 20}, 5785 (2010).
	
	\bibitem{Tanusri(2013)}
	T. Saha-Dasgupta, {\em J. Superconductivity and Novel Mag.}, {\bf 26}, 1991 (2013).
	
	\bibitem{Sami Vasala(2015)}
	S. Vasala, M. Karppinen,  {\em Progress in Solid State Chemistry}, {\bf 43}, 1 (2015).
	
	\bibitem{Tanusri(2020)}
	  T. Saha-Dasgupta, {\em Materials Research Express}, {\em 7}, 1 (2020).

\bibitem{D.D. Sarma(2000)}
	D. D. Sarma, P.  Mahadevan, T. Saha Dasgupta, S. Ray, A. Kumar,  {\em Phys. Rev. Lett.}, {\bf 85}, 2549 (2000).

\bibitem{K.-I. Kobayashi(1998)}
	K.-I. Kobayashi, T. Kimura, H. Sawada, K. Terakura, Y. Tokura, {\em  Nature (London)}, {\bf 395}, 677 (1998).
        
        \bibitem{Hena Das(2008)}
	H. Das, U. V. Waghmare, T. Saha-Dasgupta, D. D. Sarma, {\em Phys. Rev. Lett.},  {\bf 100}, 186402 (2008).

      \bibitem{DP1} A. Chattopadhyay and A. J. Millis, {\em Phys. Rev. B} {\bf 64}, 024424 (2001).

      \bibitem{DP2} L. Brey, M. J. Calderon, S. Das Sarma and F. Guinea, {\em Phys. Rev. B}, {\bf 74}, 094429 (2006).

      \bibitem{DP3} J.L.Alonso,L.A. Fernandez, F. Guinea, F. Lesmes, and V Martin-Mayor, {\em Phys. Rev. B},
        {\bf 67}, 214423 (2003).

      \bibitem{DP4} J.B. Phillip, P. Majewski, L. Alff, A. Erb, R. Gross, T.
Graf, M.S.Brandt, J. Simon, T. Walther, W. Mader, D. Topwal, and D.D. Sarma, {\em Phys. Rev.B}, {\bf 68}, 144431 (2003).

       \bibitem{Prabuddha Sanyal(2009)}
	 P. Sanyal, P. Majumdar, {\em Phys. Rev. B}, {\bf 80}, 054411 (2009).

 \bibitem{kato(2007)}	

     H. Kato et. al. {\em App. Phys. Lett}, {\bf 81}, 328 (2002).

   \bibitem{Krockenberger(2007)}	
 Y. Krockenberger et. al., {\em Phys. Rev. B}, {\bf 75}, 020404 (2007).
	
	\bibitem{Nyrissa S. Rogado(2005)}
	J. Li, A. W. Sleight, M. A. Subramanian, {\em Adv. Mater.}, {\bf 17}, 2225 (2005).
	
	\bibitem{Hena Das(2009)}
	H. Das, U. V. Waghmare, T. Saha-Dasgupta, D. D. Sarma, {\em Phys. Rev. B}, {\bf 79}, 144403 (2009).
	\bibitem{Prabuddha Sanyal(2017)}
	  P. Sanyal, {\em Phys. Rev. B}, {\bf 96}, 214407 (2017).

          	
	\bibitem{Hena Das(2011)}
	H. Das, P. Sanyal, T. Saha-Dasgupta, D. D. Sarma, {\em Phys. Rev. B}, {\bf 83}, 104418 (2011).
	\bibitem{Anita(2019)}	
	  A. Halder, P. Sanyal, T. Saha-Dasgupta, {\em Phys. Rev. B}, {\bf 99}, 020402 (R) (2019).
  \bibitem{kartik(2015)}  K Samanta, P Sanyal, T Saha-Dasgupta
Sci. Rep. {\bf 5}, 15010 (2015).
	
         
	\bibitem{E. Solana-Madruga(2016)}
	  E. Solana-Madruga, A. M. Ar\'{e}valo-L\'{o}\'{p}ez, A. J. Dos santos-Garcı\'{a}, E. Urones-Garrote,  D. Avila-Brande, R. S\'{a}\'{e}z-Puche, J. P. Attfield, {\em Angew. Chem., Int. Ed.}, {\bf 55}, 9340 (2016).
          
          	\bibitem{McNally (2017)}
	          G. M. McNally, A. M. Ar\'{e}valo-L\'{o}pez, P. Kearins, F. Orlandi, P. Manuel, J. P. Attfield, {\em Chem. Mater.}, {\bf 29}, 8870 (2017).
                  
	\bibitem{E. Solana-Madruga(2018)}
	  E. Solana-Madruga, A. M.  Ar\'{e}valo-L\'{o}pez, A. J. Dos santos-Garcı\'{a}, C. Ritter, C. Cascales, R. S\'{a}ez-Puche, J. P. Attfield, {\em Phys. Rev. B}, {\bf 97}, 134408 (2018).

   \bibitem{A. J. Dos santos-Garca(2015)}
   E. Solana-Madruga, A. J. Dos santos-Garca, A. M.  Arvalo L\'{y}pez, D. Avila-Brande, C.  Ritter,  J. P. Attfield, J. Saez-Puche, {\em R. Dalton Trans}, {\bf 44}, 20441 (2015).
	
	\bibitem {Elena Solana-Madruga(2019)}
	E. Solana-Madruga, Y.  Sun, A. M. Ar\'{e}valo-L\'{o}pez, J. P. Attfield,  {\em Chem. Commun.}, {\bf 55}, 2605 (2019).
	
	
	
	\bibitem{O. K. Andersen(2000)}
	   O. K. Andersen, T. Saha-Dasgupta, {\em Phys. Rev. B}, {\bf 62}, R16219 (2000).

\bibitem{manganites} A. J. Millis, B. I. Shraiman and R. Mueller, Phys. Rev. Lett, {\bf 77}, 175 (1996);      T. V. Ramakrishnan, H. R. Krishnamurthy, S. R. Hassan, and G. Venketeswara Pai, Phys. Rev. Lett. {\bf 92}, 157203 (2004).

 \bibitem{millis} A. Chattopadhyay and A. J. Millis, Phys. Rev. B, {\bf 64}, 024424, (2001).
           
\bibitem{Anderson-hasegawa} P.W. Anderson and H. Hasegawa, Phys. Rev. {\bf 100}, 675 (1955).


\bibitem{Gennes} P. -G de Gennes, Phys. Rev. 118, 141 (1960).  
  
      \bibitem{Goodenough} P. W. Anderson, {\em Phys. Rev.}, {\bf 79} 350 (1950); J. B. Goodenough,
        {\em Phys. Rev.} {\bf 100} 564 (1955); J. Kanamori, J. Phys. Chem. Solids. {\bf 10} 87 (1959).


	
\bibitem{das} R. Das, P. Yanda, A.Sundaresan and D.D. Sarma, Mater. Res. Express6 116122 (2019).
	
\bibitem{nicro} A. Ali, G. Sharma, Y. Singh, arXiv:1811.07836.
        
\bibitem{G. Kresse(1996)}
	G. Kresse, J. Furthmller, {\em Computational Materials
	  Science}, {\bf 6(1)}, 15 (1996).
        	
	
	\bibitem{PBE} J. P. Perdew, K. Burke, M. Ernzerhof, {\em  Phys. Rev. Lett.}, {\bf 77}, 3865 (1996); ibid, {\bf 78}, 1396(E) (1991).
        
	
	\bibitem{P. E. Blochl(1994)}
	  P. E.   Bl\"{o}chl, {\em Phys. Rev. B}, {\bf 50}, 17953 (1994).
          
	\bibitem{S. L. Dudarev(1998)}
	S. L. Dudarev, G. A. Botton, S. Y. Savrasov, C. J. Humphreys, A. P. Sutton, {\em Phys. Rev. B}, {\bf 57(3)}, 1505 (1998). 
        
\bibitem{doublecounting} E. R. Ylvisaker, W. E. Pickett, and K. Koepernik,
{\em Phys. Rev. B} {\bf 79}, 035103 (2009).
        
        \bibitem{U-value} I. V.  Solovyev, P. H. Dederichs, V. I. Anisimov,
{\em Phys. Rev. B}, {\bf 50}, 16861 (1994).
	
\bibitem{lmto} O. K. Andersen, O. Jepsen, {\em Phys. Rev. Lett.}, {\bf 53}, 2571 (1994).

  
\end{thebibliography}
\end{document}